\documentclass[twocolumn,aps,pre,preprintnumbers,amsmath,amssymb]{revtex4-1}
\usepackage{graphicx}				
\usepackage{dcolumn}				
\usepackage{bm}						

\begin{document}
\title{Can Local Stress Enhancement Induce Stability in Fracture Processes? Part II: The Shielding Effect}
\author{Jonas T. Kjellstadli}
\email{jonas.kjellstadli@outlook.com}
\author{Eivind Bering}
\email{eivind.bering@ntnu.no}
\author{Srutarshi Pradhan}
\email{srutarshi.pradhan@ntnu.no}
\author{Alex Hansen}
\email{alex.hansen@ntnu.no}
\affiliation{PoreLab, Department of Physics, NTNU -- Norwegian University of Science and Technology, Trondheim, Norway}

\date{\today}

\begin{abstract}
We use the local load sharing fiber bundle model to demonstrate a shielding effect where strong fibers protect weaker ones. This effect exists due to the local stress enhancement around broken fibers in the local load sharing model, and it is therefore not present in the equal load sharing model. The shielding effect is prominent only after the initial disorder-driven part of the fracture process has finished, and if the fiber bundle has not reached catastrophic failure by this point, then the shielding increases the critical damage of the system, compared to equal load sharing. In this sense, the local stress enhancement may make the fracture process more stable, but at the cost of reduced critical force.
\end{abstract}

\maketitle

\section{Introduction}
\label{intro}

When brittle materials fail mechanically under loading, the failure is the end point of a competition between local stress and local strength in the material. They pull in opposite directions \cite{rh90}. When there is a local failure somewhere in the material, stresses increase at that location, which increases the likelihood that the subsequent failure happens in that neighborhood. One may say that stresses make local failures attract each other. Disorder in the strength of the material on the other hand, has the opposite effect. This is a purely statistical effect: the further away from the failure, the weaker the weakest spot in the material will be. Hence, the disorder drives local failures apart; they induce repulsion between the local failures.

\begin{figure}
\begin{center}
\includegraphics[width=\columnwidth]{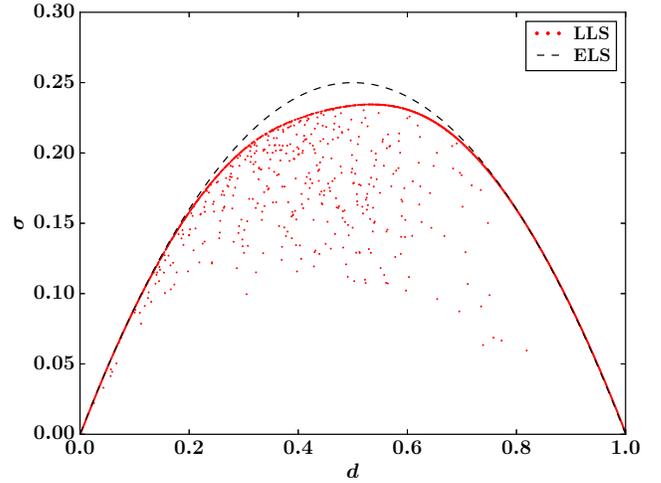}
\caption{Stress $\sigma$ vs.\ damage $d$ in the equal load sharing (ELS) fiber bundle model and the local load sharing (LLS) fiber bundle model on a square lattice. The ELS curve has been calculated analytically via equation (\ref{eq:load_curve_ELS}), and the LLS result is a simulation of a single sample ($N = 2048^{2}$). For clarity, the LLS result shows only every 2500th data point. The threshold distribution is uniform on the unit interval for both models.}
\label{fig1}
\end{center}
\end{figure}

When the damaged zones grow, the stresses at their edges increase and at some point, the repulsion induced by the disorder in the local strength is no longer able to counter this effect. When this occurs, catastrophic failure ensues. But this picture is not the whole story. We show in Figure \ref{fig1} the stress $\sigma$ as a function of the damage $d$ for two fracture models: The equal load sharing (ELS) fiber bundle model (FBM) and the local load sharing (LLS) FBM \cite{phc10,hhp15}, to be described in Section \ref{fbm}. The LLS model contains stress enhancement at the edge of the damaged zones, i.e., clusters of broken fibers, whereas the ELS model does not. As expected we see that the ELS model is stronger than the LLS model since the maximum value of $\sigma$ is larger for this model than for the LLS model. However, one curious feature stands out in this figure: \textit{The LLS model reaches its maximum value of $\sigma$ for a higher damage $d$ than the ELS model.} In other words, the LLS model where there is stress enhancement may sustain higher damage than the ELS model where there is no stress enhancement. We will show that this effect is due to \textit{shielding} \cite{b16} of weak areas by strong areas.

We discussed in Part I \cite{kbhph19} a different mechanism that would lead to an \textit{apparent} stability of the LLS model when the ELS model is unstable. This turned out to be a purely statistical effect coming from averaging over many samples. In Figure \ref{fig1}, we show only a single sample for the LLS model; there is no averaging. Hence, the shielding effect is a real effect that can be observed in single samples.

In Section \ref{fbm} we describe the ELS and LLS fiber bundle models. We then go on in Section \ref{shielding} to give a detailed explanation of the shielding effect in terms of the LLS model in one dimension. In Section \ref{evidence} we demonstrate that the shielding effect is sufficiently common to produce the effect already seen in Figure \ref{fig1}. In Section \ref{effect} we demonstrate two effects in the LLS model compared to the ELS model that we attribute to the shielding effect, and Section \ref{relevancy} deals with determining when the shielding effect is relevant. We end by summarizing and discussing our results in Section \ref{discussion}.

\section{The Fiber Bundle Model}
\label{fbm}

We start by defining the Fiber Bundle Model \cite{phc10,hhp15}. $N$ elastic fibers with identical spring constants $\kappa$ are placed between two clamps. In the simulations and results we will use $\kappa = 1$ for simplicity. The only effect of changing $\kappa$ is to rescale the forces. A fiber $i$ acts like Hookean spring, the force it carries given by $f_{i} = \kappa x$, for an elongation $x$ smaller than its threshold $t_{i}$, which is individual for each fiber. When the elongation reaches the threshold, the fiber breaks irreversibly and cannot carry a force anymore. The thresholds are drawn from a probability distribution, denoted by the cumulative distribution $P(t)$, which is a parameter of the model. The number of broken fibers is denoted by $k$. Thus $k/N$ is the fraction of broken fibers, also called the damage $d$. We use quasistatic loading of the model, where the load is increased until it is sufficient to break a single fiber, and then immediately lowered.

\subsection{Equal Load Sharing}
\label{els}

To determine what happens when fibers break, a load sharing rule is required. The simplest one is the ELS scheme \cite{p26,d45}, also known as global load sharing, where every intact fiber shares the applied load equally. It corresponds to the clamps being infinitely stiff (as long as the spring constant $\kappa$ is identical for the fibers). This means that there is no stress enhancement around the fibers that fail. A further consequence is that the fibers break in order of increasing thresholds as the applied force is increased, regardless of whether quasistatic loading is used.

With $k = NP(x)$ broken fibers at elongation $x$ \cite{g04} and a total external force $F = N \sigma$, ELS results in the relation
\begin{equation}
\sigma = \kappa x \left( 1 - P(x) \right) = \kappa P^{-1}(d) \left( 1 - d \right)
\label{eq:load_curve_ELS}
\end{equation}
between the force per fiber $\sigma$ required to break the next fiber and the elongation $x$ (or the damage $d$) of the fiber bundle. Note that equation (\ref{eq:load_curve_ELS}) is exact in the limit $N \rightarrow \infty$, but for finite system sizes there are fluctuations around this average behavior \cite{hhp15}.

For nearly all choices of threshold distribution $P(t)$, equation (\ref{eq:load_curve_ELS}) has a single maximum $\sigma_{c}$, the critical strength of the bundle, at which the fiber bundle collapses. There is a corresponding critical elongation $x_{c}$ and critical damage $d_{c} = P(x_{c})$. These quantities are also defined for local load sharing, which we will discuss next, but there they are not available from simple analytical expressions.

\subsection{Local Load Sharing}
\label{lls}

A different load sharing rule is the LLS one \cite{hp78}, where the extra load from broken fibers is distributed equally onto the nearest intact neighbors in the lattice the fibers are placed on. As a consequence, LLS behaves differently when the underlying lattice changes. This is different from ELS, where fiber placement is irrelevant, since the load is assumed to be evenly distributed.

For LLS one must choose a lattice for fiber placement. A \textit{hole} is then defined as a cluster of $h$ broken fibers connected by nearest neighbor connections. The \textit{perimeter} of a hole is the $p$ intact fibers that are nearest neighbours of the hole. With an applied force per fiber $\sigma = F/N$, the force acting on an intact fiber $i$ can then be expressed as
\begin{equation}
f_{i} = \sigma \left( 1 + \sum_{j} \frac{h_{j}}{p_{j}} \right).
\label{eq:load_curve_LLS}
\end{equation}
Here $j$ runs over holes neighboring the fiber. The two terms can be interpreted as respectively the force originally applied to every fiber, and the redistribution of forces due to broken fibers.

LLS was originally defined for a one-dimensional lattice with periodic boundary conditions \cite{hp78}, but the formulation in equation (\ref{eq:load_curve_LLS}) is a generalization applicable to any lattice. We mainly study LLS on a two-dimensional square lattice in this paper. We also use periodic boundary conditions for the lattices we study.

Equation (\ref{eq:load_curve_LLS}) is \textit{history independent}: the breaking order of fibers does not affect the load redistribution. This is the way LLS was defined originally \cite{hp78}, but some later implementations have been history dependent, where the load a fiber carries is simply divided among its nearest neighbors when it breaks \cite{kzh00} making it impossible to determine the load a fiber carries without knowing the order in which the fibers up to that point have failed. In 1D this approach does not give very different results from the history independent model, since 1D LLS has zero critical damage and collapses due to extreme loads on fibers that neighbor large holes. However, in dimensions $D > 1$, the history dependent model gives very different results \cite{dok18} from the history independent model \cite{skh15}.

To determine a failure criterion for individual fibers we define the \textit{effective threshold} $t_{\text{eff},i}$ of fiber $i$ as
\begin{equation}
t_{\text{eff},i} = \frac{t_{i}}{1 + \sum_{j} \frac{h_{j}}{p_{j}}}.
\label{eq:effective_threshold_LLS}
\end{equation}
The effective thresholds depend both on the original thresholds $t_{i}$ of the fibers and the hole structure of the bundle, meaning that they change as the fiber bundle breaks down. By combining this expression with equation (\ref{eq:load_curve_LLS}) we find the breaking criterion $\sigma = \kappa t_{\text{eff},i}$ where the fiber with the smallest effective threshold fails under the smallest external load $\sigma$.

Hence quasistatic loading results in a fracture process where the next fiber to break is always the one with the smallest effective threshold, given by equation (\ref{eq:effective_threshold_LLS}). When a fiber is broken, effective thresholds must be updated to determine which fiber breaks next.

\section{Defining Shielding}
\label{shielding}

Let us investigate a simple example in 1D to demonstrate what we mean by shielding. Consider $N = 10$ fibers with thresholds $\{t_{i}\} = \{0.1,0.2,...,1.0\}$ arranged as follows:
\begin{equation}
1.0 \quad 0.1 \quad 0.2 \quad 0.9 \quad 0.8 \quad 0.3 \quad 0.7 \quad 0.5 \quad 0.6 \quad 0.4.
\label{eq:1D_example}
\end{equation}
With ELS we find a critical strength $\sigma_{c,\text{ELS}} = 0.3 \kappa$, but what about LLS?

When all fibers are intact the effective thresholds are identical to the original thresholds in equation (\ref{eq:1D_example}). The first fiber to break with LLS is the one with threshold $t = 0.1$, which happens at $\sigma = 0.1 \kappa$. If we let $\times$ represent a broken fiber, then the effective thresholds after breaking the first fiber are
\begin{equation*}
\frac{2}{3} \quad \times \quad \frac{2}{15} \quad 0.9 \quad 0.8 \quad 0.3 \quad 0.7 \quad 0.5 \quad 0.6 \quad 0.4.
\end{equation*}
The broken fiber constitutes a hole of size $h = 1$, with $p = 2$ fibers in its perimeter. These two fibers are the only ones whose effective thresholds change when the first fiber breaks. From equation (\ref{eq:effective_threshold_LLS}) we see that their new effective thresholds are their original thresholds divided by $1 + 1/2$.

The effectively weakest fiber now breaks at $\sigma = 2\kappa/15$, and the effective thresholds afterward are
\begin{equation*}
0.5 \quad \times \quad \times \quad 0.45 \quad 0.8 \quad 0.3 \quad 0.7 \quad 0.5 \quad 0.6 \quad 0.4,
\end{equation*}
since there is a single hole with $h = p = 2$. The third fiber breaks at $\sigma = 0.3 \kappa = \sigma_{c,\text{ELS}}$, which results in the effective thresholds
\begin{equation*}
0.5 \quad \times \quad \times \quad 0.45 \quad \frac{8}{15} \quad \times \quad \frac{7}{15} \quad 0.5 \quad 0.6 \quad 0.4.
\end{equation*}
The smallest effective threshold is $0.4$, and once this fiber breaks, the fiber bundle collapses. Hence the critical strength is $\sigma_{c,\text{LLS}} = 0.4 \kappa > \sigma_{c,\text{ELS}}$.

With ELS, $0.3 \kappa$ is the critical strength because the fibers with thresholds $0.4$ and $0.5$ receive some of the redistributed load from the broken fibers. They break at $\sigma = 0.28 \kappa$ and $\sigma = 0.3 \kappa$, respectively.

With LLS, the four strongest fibers happen to neighbor the three broken ones and receive all redistributed load. In this sense, the other three intact fibers (the three rightmost ones) are \textit{shielded} from this additional load, and their effective thresholds are unchanged from the fully intact fiber bundle. In our example the result is an increased critical strength and critical damage compared to ELS, because the four strong fibers that receive additional loads don't have their effective thresholds lowered below $0.3$.

But this example is contrived. A large system ($N \rightarrow \infty$) \textit{will} contain strong configurations like the one in our example, but it will also contain weak configurations with many adjacent fibers that all have small thresholds. In 1D, a hole can never have a larger perimeter than $p = 2$. Therefore a sufficiently large hole (that originates at a particularly weak configuration) will reduce the effective thresholds of its neighboring fibers enough that they also break, inducing a fatal rupture that opens the fiber bundle like a ziplock. Strong configurations, where shielding is important, are ripped open by this expanding hole.

The result is that in 1D, as $N \rightarrow \infty$, the critical damage of LLS goes to zero and the critical strength goes to the lower limit of the threshold distribution \cite{hhp15}. Both of which are much smaller than their corresponding ELS values.

Still, our example highlights an interesting effect: with localized force distribution, strong fibers can shield weaker ones from some of the applied load. The question of interest is whether there is a noticeable shielding effect in the LLS model on lattices in $D > 1$, and, if so, with what consequences?

\section{Evidence of Shielding}
\label{evidence}

Let $w$ be the intact fiber with the smallest threshold, i.e., the weakest intact fiber. We then study be the probability $p_{w}$ that $w$ is the first fiber to break when the applied load $\sigma$ is increased.

With ELS we get $p_{w} = 1$, since all intact fibers share the same load. With LLS this is not the case, because the fiber with the smallest \textit{effective} threshold breaks. Equation (\ref{eq:effective_threshold_LLS}) shows that a small effective threshold results from a combination of small threshold and large force redistribution.

\begin{figure}
\begin{center}
\includegraphics[width=\columnwidth]{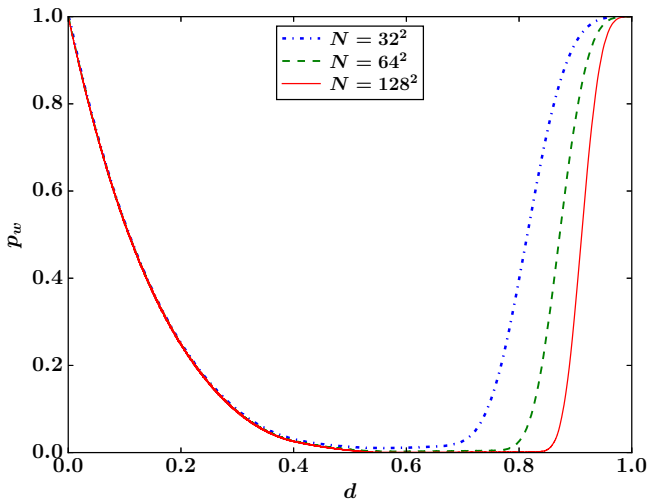}
\caption{Probability $p_{w}$ to break the intact fiber with the smallest threshold in LLS simulations on a square lattice. The threshold distribution is uniform: $P(t) = t$. The number of samples is $8 \times 10^{5}$, $2 \times 10^{5}$, and $10^{5}$ for system sizes $N = 32^{2}$, $64^{2}$, and $128^{2}$, respectively.}
\label{fig2}
\end{center}
\end{figure}

Figure \ref{fig2} shows $p_{w}$ as a function of the damage $d$ for the LLS model with $P(t) = t$ on a square lattice for different system sizes $N$. Throughout most of the fracture process $p_{w}$ is small, i.e., it is unlikely that $w$ will break at any given step. This indicates that load redistribution dominates the effective thresholds, and that at least some of the fibers with small thresholds are partially shielded from the applied load.

There are significant finite size effects for $p_{w}$ in Figure \ref{fig2}. Finite size scaling indicates that in the thermodynamic limit $N \rightarrow \infty$, there is a sharp transition from $p_{w} = 0$ to $p_{w} = 1$ around $d \approx 0.98$. When all intact fibers neighbor a single hole, then $p_{w} = 1$, and the damage at which the transition happens should therefore be expected to change with the lattice.

To study shielding further we define a \textit{load sharing factor} $\sigma/f_{i}$, the ratio between the applied load $\sigma$ and the force $f_{i}$ acting on fiber $i$. For ELS the force $f_{i,\text{ELS}}$ is identical for all intact fibers, and there is an exact expression
\begin{equation}
\frac{\sigma}{f_{i,\text{ELS}}} = 1 - d
\label{eq:load_sharing_factor_ELS}
\end{equation}
in the limit $N \rightarrow \infty$. For LLS the load sharing factor depends on which fibers we follow through the breaking process.

\begin{figure}
\begin{center}
\includegraphics[width=\columnwidth]{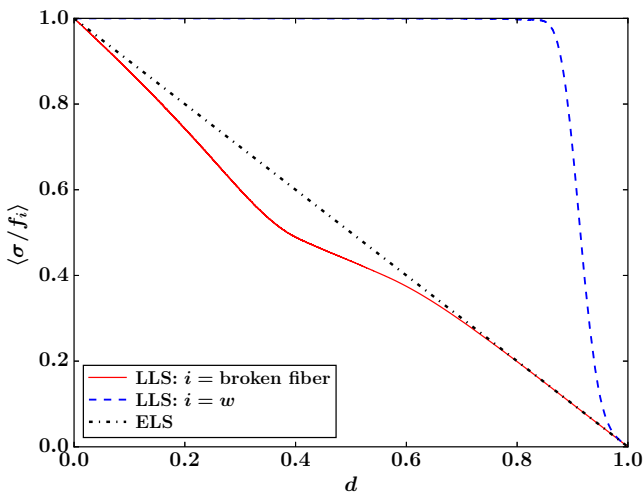}
\caption{Sample averaged load sharing factor $\langle \sigma/f_{i} \rangle$ for LLS with $i = w$ and $i$ being the broken fiber, and for ELS (equation (\ref{eq:load_sharing_factor_ELS})). The threshold distribution is uniform ($P(t) = t$) and the LLS results are from simulations on a square lattice ($N = 128^{2}$) averaged over $10^{5}$ samples.}
\label{fig3}
\end{center}
\end{figure}

Figure \ref{fig3} contrasts the load sharing factor for ELS and LLS. The broken fibers in the LLS model are on average more loaded than in the ELS model, which is expected; highly loaded fibers are more likely to break since their effective thresholds are reduced significantly from the original thresholds.

It is more interesting that $w$, the intact fiber with the smallest threshold, on average recieves almost no extra load from redistribution throughout most of the breaking process. In particular, the average fraction of the load it receives is much smaller than fibers in the ELS model. While one cannot in general trust averages in the LLS model blindly \cite{kbhph19}, this does indicate that these weak fibers are shielded from some of the applied load, and that fibers with higher thresholds (but smaller effective thresholds due to being highly loaded) break instead of them.

That the average load sharing factor of the fiber $w$ decreases rapidly from $1$ (Figure \ref{fig3}) around the same damage that $p_{w}$ increases quickly (Figure \ref{fig2}) is not a coincidence. When the intact fiber with the smallest threshold becomes more highly loaded, it is very likely to have the smallest effective threshold, and hence the probability that it breaks increases. The finite size effects of the load sharing factor in Figure \ref{fig3} for $w$ therefore mirror the ones for $p_{w}$ in Figure \ref{fig2}.

\section{Effects of Shielding}
\label{effect}

\begin{figure}
\begin{center}
\includegraphics[width=\columnwidth]{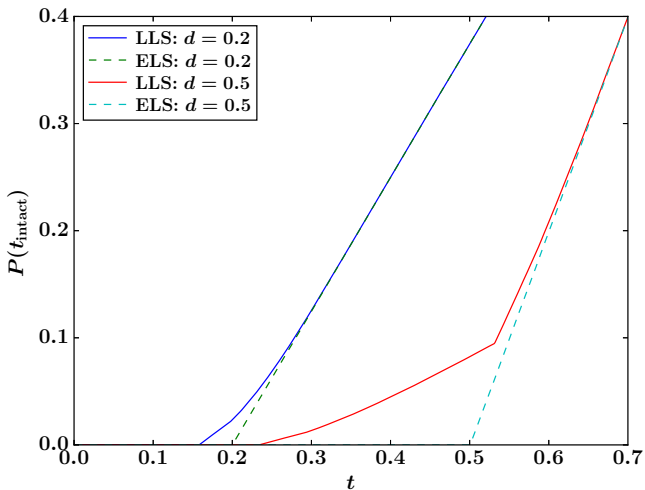}
\caption{Cumulative distribution $P(t_{\text{intact}})$ of fiber thresholds of intact fibers at two different damages $d$ for a single sample ($N = 1024^2$). LLS results are from a simulation on a square lattice, ELS results are calculated using the same thresholds as in the LLS simulation. The threshold distribution of all fibers is uniform: $P(t) = t$.}
\label{fig4}
\end{center}
\end{figure}

Since the shielding effect protects the weakest intact fibers, we intuitively expect that LLS will have more intact weak fibers than ELS. To investigate this hypothesis, Figure \ref{fig4} shows the cumulative probability distribution $P(t_{\text{intact}})$ of fiber thresholds of intact fibers for a single sample with $N = 1024^{2}$ fibers. The thresholds were drawn from a uniform distribution $P(t) = t$.

With a damage of $d = 0.2$, there are some intact fibers in LLS that have smaller threshold than the intact fibers in ELS, but the difference between the two load sharing rules is not large. This is because the LLS model behaves similarly to ELS in the early stages of the breaking process. The disorder of the threshold distribution dominates. Hence fibers fail because they have small thresholds, rather than because they are highly loaded \cite{skh15}. In this regime there is little room for the shielding, which is an effect of the LLS rule, to affect the fiber bundle significantly.

This changes when the damage increases, as Figure \ref{fig2} shows. When $k/N = 0.5$, slightly below the critical damage of the sample for both LLS and ELS, there is a significant difference between the threshold distributions of intact fibers for the two load sharing rules. With LLS, the weakest intact fibers have thresholds $t < 0.25$, while with ELS the lower limit for thresholds is $t = 0.5$. Approximately $8.2\%$ of intact fibers in LLS have thresholds smaller than the lower limit for ELS. Thus the shielding effect that emerges from LLS protects some fibers with small thresholds, which survive longer than they would have with ELS.

Note that only a small fraction of the weak fibers are shielded by this effect. Out of the $523,690$ fibers with thresholds smaller than $0.5$ in the analyzed sample, only approximately $8.2\%$ are intact at damage $d = 0.5$.

\begin{figure}
\begin{center}
\includegraphics[width=\columnwidth]{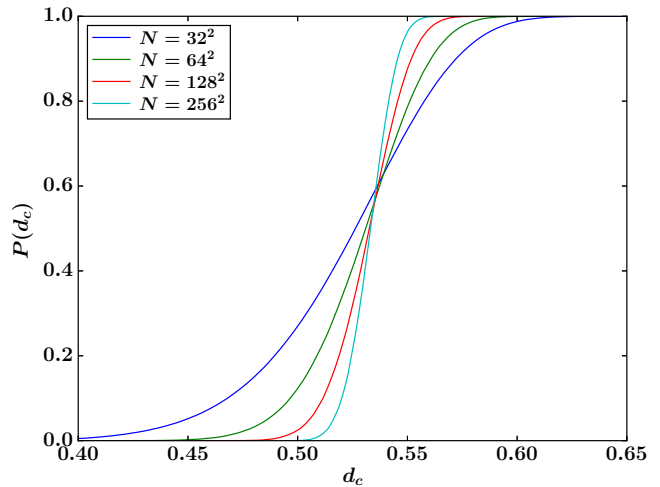}
\caption{Cumulative distribution of critical damage $P(d_{c})$ for LLS on a square lattice. The number of samples is $8 \times 10^{5}$, $2 \times 10^{5}$, $10^{5}$, and $3 \times 10^{4}$ for system sizes $N = 32^{2}$, $64^{2}$, $128^{2}$, and $256^{2}$, respectively. The threshold distribution is uniform: $P(t) = t$.}
\label{fig5}
\end{center}
\end{figure}

Another effect of shielding is an increase in the critical damage $d_{c}$, the fraction of fibers broken when the fiber bundle collapses, compared to ELS. For a uniform threshold distribution $P(t) = t$, ELS has $d_{c} = 1/2$. Figure \ref{fig5} shows the cumulative probability distribution of $d_{c}$ for LLS on a square lattice, also for a uniform threshold distribution. As $N \rightarrow \infty$, the critical damage converges to a value that is roughly $5\%$ larger than the ELS value.

This means that shielding has the surprising effect of making LLS \textit{more stable} than ELS. An LLS fiber bundle reaches catastrophic failure at a higher damage than a corresponding ELS fiber bundle, and there is a region with $d$ slightly larger than $1/2$ where ELS is unstable (it has passed the greatest force it can sustain before breaking), while LLS is not (it has yet to reach this point).

\begin{figure}
\begin{center}
\includegraphics[width=\columnwidth]{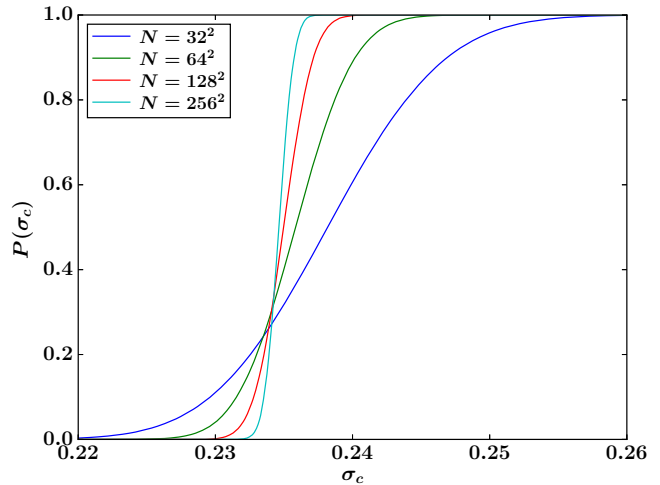}
\caption{Cumulative distribution of critical strength $P(\sigma_{c})$ for LLS on a square lattice. The number of samples is $8 \times 10^{5}$, $2 \times 10^{5}$, $10^{5}$, and $3 \times 10^{4}$ for system sizes $N = 32^{2}$, $64^{2}$, $128^{2}$, and $256^{2}$, respectively. The threshold distribution is uniform: $P(t) = t$.}
\label{fig6}
\end{center}
\end{figure}

However, this increased stability comes at the cost of a reduced critical strength $\sigma_{c}$. ELS has $\sigma_{c} = 1/4$ for $P(t) = t$, and Figure \ref{fig6} shows the corresponding values for LLS on a square lattice. As $N \rightarrow \infty$ the LLS critical strength converges toward $\sigma_{c} \approx 0.233$, approximately $7\%$ smaller than the ELS value.

\begin{figure}
\begin{center}
\includegraphics[width=\columnwidth]{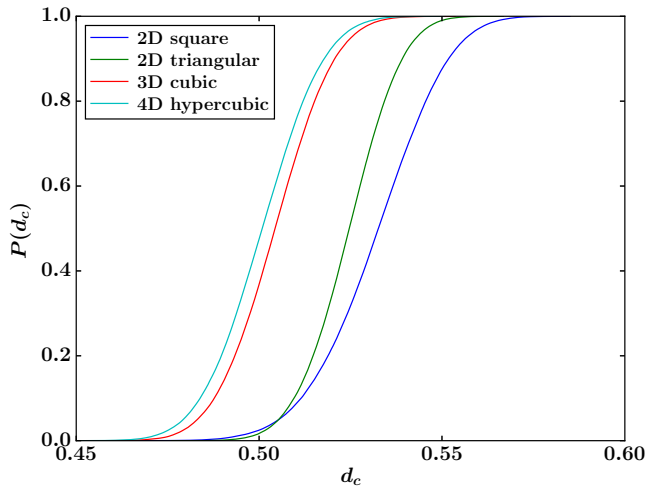}
\caption{Cumulative distribution of critical damage $P(d_{c})$ for LLS on four lattices with similar number of fibers: square ($N = 128^{2}$), triangular ($N = 128^{2}$), cubic ($N = 25^{3}$), and 4D hypercubic ($N = 11^{4}$), all simulated with $10^{5}$ samples. The threshold distribution is uniform: $P(t) = t$.}
\label{fig7}
\end{center}
\end{figure}

So far we have investigated LLS on a square lattice only, but effects of the proposed shielding mechanism should be present in all other lattices (except for 1D, as explained earlier). Nothing about the proposed shielding effect is specific to the square lattice, but we should expect that the effects become smaller as the connectivity of the lattice increases; it is less likely that a weak fiber is surrounded (and hence shielded) by strong fibers in a higher-dimensional lattice or a lattice with higher connectivity. To test this hypothesis, Figure \ref{fig7} shows the distribution of critical damage for LLS on four different lattices.

As expected, the critical damage is highest for the square lattice, because the shielding effect is the most pronounced there. The square lattice has the lowest connectivity and dimension of the four lattices that is tested here.

The triangular lattice has a smaller critical damage, but still significantly greater than ELS. This is consistent with a somewhat less pronounced shielding effect, which is expected for a lattice that is also two-dimensional, but with higher connectivity than the square lattice.

The cubic and 4D hypercubic lattices have much smaller critical damages than the two-dimensional lattices, and are comparable to ELS. This is consistent with the proposed shielding effect, which should become much weaker as the dimension increases; there are many more possible paths for a hole to gain access to the inside of a shielded region in higher dimensions, and shielding is therefore much rarer.

\section{When Is Shielding Relevant?}
\label{relevancy}
Early in the breaking process the LLS model behaves ELS-like, as corroborated by Figures \ref{fig1}, \ref{fig2}, \ref{fig3}, and \ref{fig4}. The disorder of the threshold distribution dominates the process, i.e., the effective thresholds from equation (\ref{eq:effective_threshold_LLS}) are dominated by the original thresholds in the numerator, not the hole structure in the denominator. What happens when the fiber bundle has a critical damage in this disorder-dominated regime? If the increased critical damage of LLS compared to ELS for the uniform distribution is indeed an effect of shielding, one would expect that for threshold distributions where the critical damage $d_{c}$ is in the disorder-dominated regime, LLS has a smaller critical damage than ELS due to the local stress enhancement.

\begin{figure}
\begin{center}
\includegraphics[width=\columnwidth]{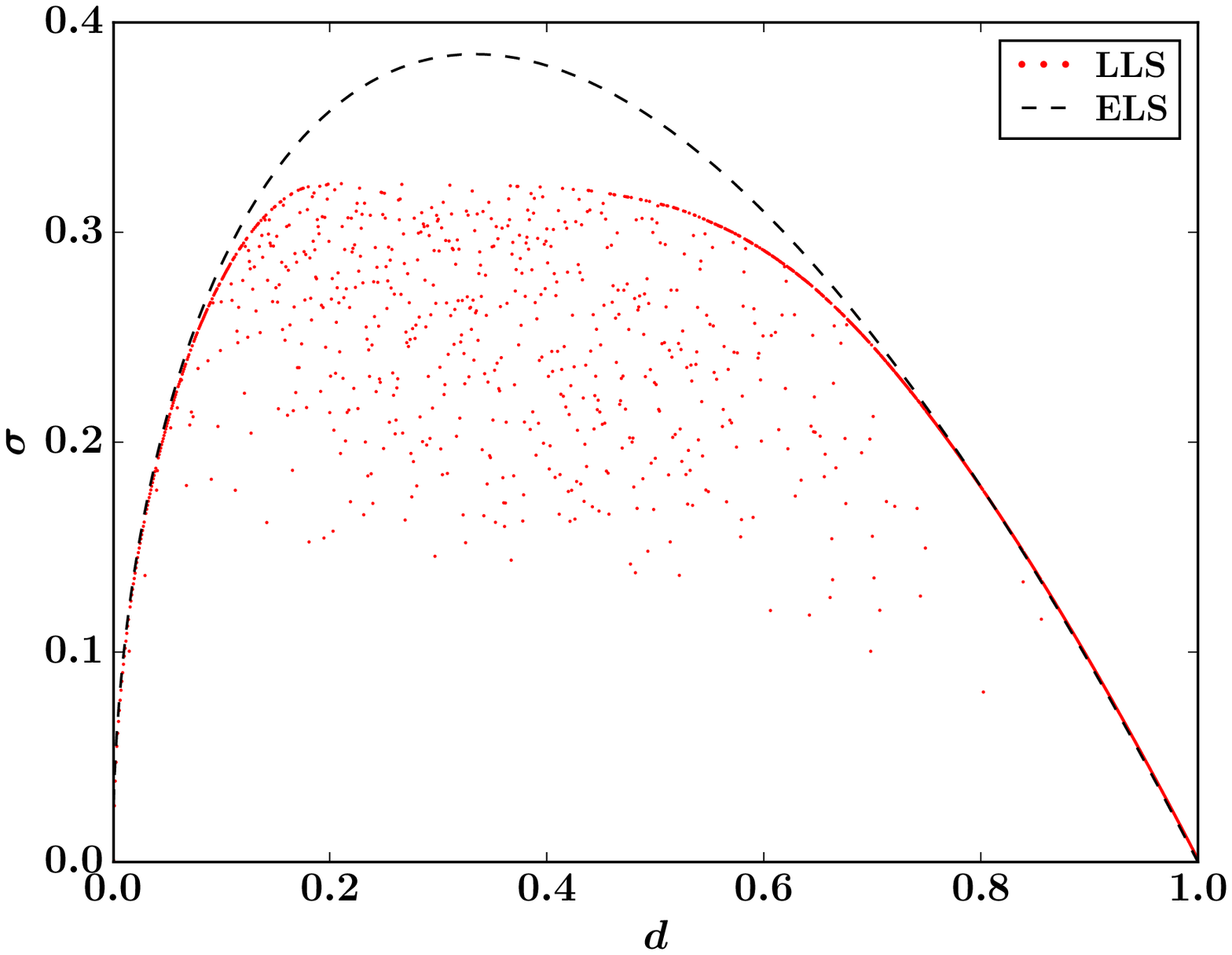}
\caption{Stress $\sigma$ vs.\ damage $d$ in the ELS model and the LLS model on a square lattice. The ELS curve has been calculated analytically via equation (\ref{eq:load_curve_ELS}), and the LLS result is a simulation of a single sample ($N = 1024^{2}$). For clarity, the LLS result shows only every 800th data point. The threshold distribution is $P(t) = t^{2}$ for both models.}
\label{fig8}
\end{center}
\end{figure}

To demonstrate this, we choose the threshold distribution $P(t) = t^{2}$. The ELS model then has a critical damage $d_{c} = 1/3$, close to the disorder-dominated regime. What about the LLS model? Figure \ref{fig8} compares the stress $\sigma$ vs.\ damage $d$ for the ELS model and the LLS model on a square lattice. As in Figure \ref{fig1} for the uniform distribution, the LLS model has a smaller critical strength $\sigma_{c}$ than the ELS model, but in this case the critical damage $d_{c}$ also seems to be smaller, approximately $d_{c} \approx 0.2$.

\begin{figure}
\begin{center}
\includegraphics[width=\columnwidth]{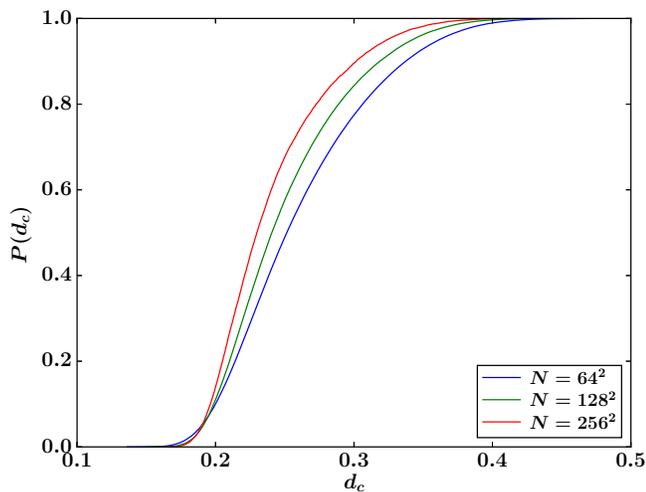}
\caption{Cumulative distribution of critical damage $P(d_{c})$ for LLS on a square lattice. The number of samples is $10^{6}$, $1.5 \times 10^{5}$, and $1.5 \times 10^{4}$ for system sizes $N = 64^{2}$, $128^{2}$, and $256^{2}$, respectively. The threshold distribution is $P(t) = t^{2}$.}
\label{fig9}
\end{center}
\end{figure}

To determine the critical damage of the LLS model more accurately we plot its cumulative probability distribution for different system sizes in Figure \ref{fig9}, like we did in Figure \ref{fig5} for the uniform distribution. It indicates that, in the limit $N \rightarrow \infty$, the critical damage is $d_{c} < 0.2$, i.e., much smaller than for the ELS model.

This is consistent with the expanation that the shielding effect is responsible for the increased critical damage of the LLS model for the uniform threshold distribution. When catastrophic failure occurs early in the breaking process, i.e., in the disorder-dominated regime, the shielding effect hardly influences the fiber bundle in the stable phase; shielding is stronger the more the hole structure of the fiber bundle dominates the effective thresholds, and it is therefore weak in the disorder-dominated regime, as corroborated by the results for $d = 0.2$ in Figure \ref{fig4}. Hence, for this kind of threshold distribution, the local stress enhancement of LLS leads to decreases in both critical strength $\sigma_{c}$ and critical damage $d_{c}$ compared to the ELS model, since shielding is not relevant in the stable phase. However, if catastrophic failure occurs late enough for the shielding effect to be relevant in the stable phase --- exactly what damages count as ``late enough'' will depend on the lattice --- then it leads to the LLS model having a higher critical damage than the ELS model.

\begin{figure}
\begin{center}
\includegraphics[width=\columnwidth]{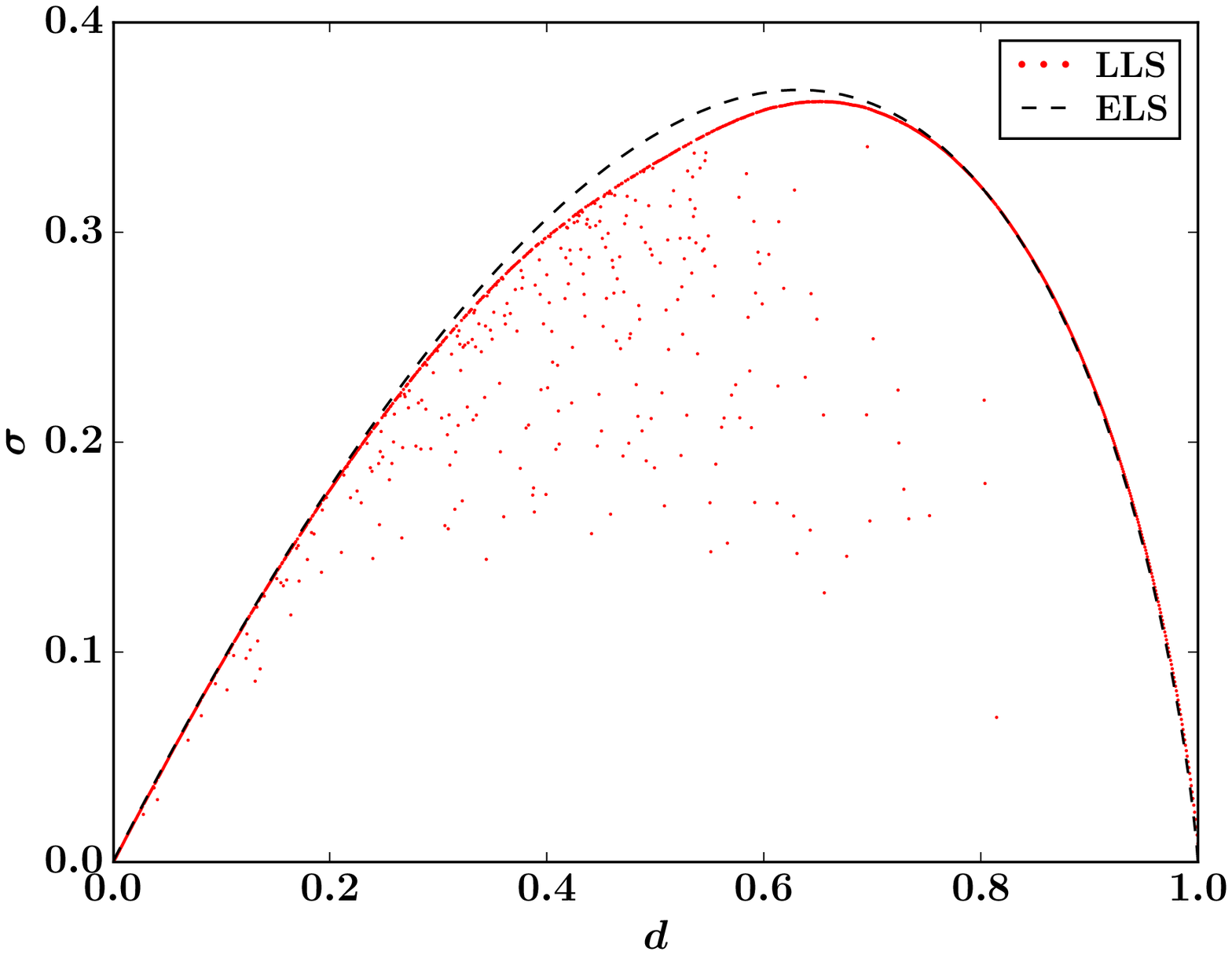}
\caption{Stress $\sigma$ vs.\ damage $d$ in the ELS model and the LLS model on a square lattice. The ELS curve has been calculated analytically via equation (\ref{eq:load_curve_ELS}), and the LLS result is a simulation of a single sample ($N = 1024^{2}$). For clarity, the LLS result shows only every 800th data point. The threshold distribution is $P(t) = 1 - \text{e}^{-t}$ for both models.}
\label{fig10}
\end{center}
\end{figure}

\begin{figure}
\begin{center}
\includegraphics[width=\columnwidth]{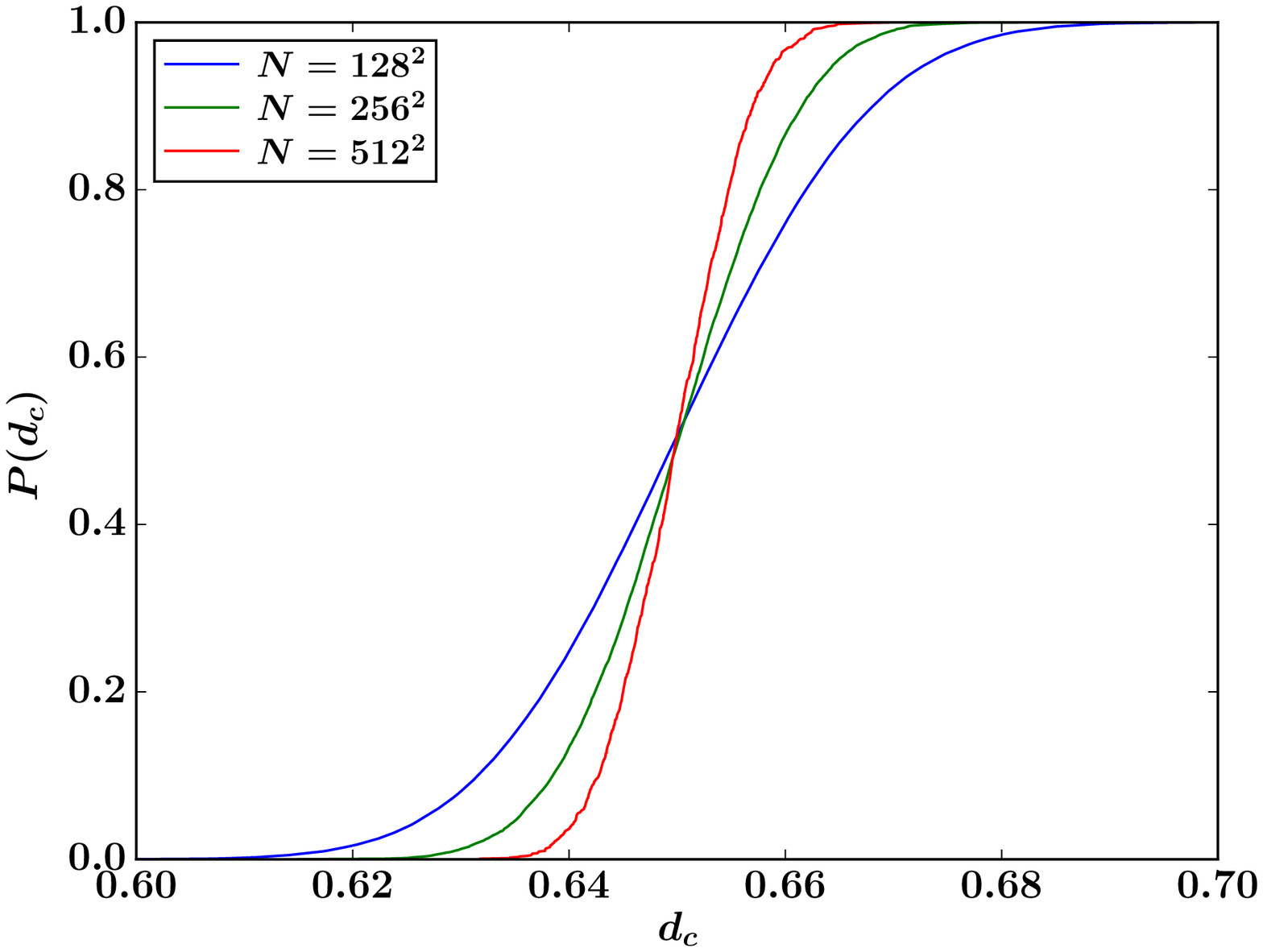}
\caption{Cumulative distribution of critical damage $P(d_{c})$ for LLS on a square lattice. The number of samples is $1.5 \times 10^{5}$, $1.5 \times 10^{4}$, and $1.5 \times 10^{3}$ for system sizes $N = 128^{2}$, $256^{2}$, and $512^{2}$, respectively. The threshold distribution is $P(t) = 1 - \text{e}^{-t}$.}
\label{fig11}
\end{center}
\end{figure}

An example of a threshold distribution where the shielding effect is relevant is the Weibull distribution $P(t) = 1 - \text{e}^{-t}$. Figure \ref{fig10} shows the stress $\sigma$ vs.\ the damage $d$ for this threshold distribution with LLS on a square lattice and ELS. For the ELS model, $d_{c} = 1 - \text{e}^{-1} \approx 0.632$, which is smaller than the critical damage for the LLS model, as evidenced by the cumulative distributions of the critical damage in Figure \ref{fig11}.


\section{Discussion}
\label{discussion}

We have shown that the LLS fiber bundle model contains a shielding effect where some of the fibers with the smallest thresholds (i.e., the weakest fibers) among the intact fibers are shielded from some of the applied load, compared to ELS. Increased connectivity and dimension of the lattice makes shielding less probable, and hence the effects decrease as the dimension or connectivity increases. The exception to this behavior is that the effect is not noticeable in 1D, where LLS has zero critical damage and strength. It is not clear if the shielding effect is important for applications where a three-dimensional model is appropriate. But for cases where a two-dimensional model is correct, the shielding effect can be expected to give important contributions to the behavior of the fracture process.

Shielding has two major effects. The first is that two-dimensional LLS models can be more stable than corresponding ELS models, in the sense that catastrophic failure occurs at a higher damage. This is, however, accompanied by a reduced critical strength. In total, LLS can, surprisingly, be preferable to ELS in two dimensions when stability is more important than strength for the application in question.

The second effect is that weak fibers are better protected and survive longer in LLS than in ELS. This is in some ways similar to how cars are built to protect the people inside at the expense of the sturdiness of the car itself. Potential applications where it is more important to have weak fibers survive than that the strength of the entire fiber bundle is high will be better off using LLS instead of ELS.

We have mainly studied the uniform threshold distribution $P(t) = t$, but also shown that shielding is an important effect for other threshold distributions --- like the Weibull distribution $P(t) = 1 - \text{e}^{-t}$ --- where catastrophic failure occurs after the initial disorder-dominated regime. However, if failure does occur in the disorder-dominated regime, before the shielding effect is strong enough, shielding will not give significant contributions to the behavior in the stable phase of the fracture process. We therefore expect that the shielding effect is universal in the sense that for the class of threshold distributions where catastrophic failure happens sufficiently late, the shielding effect will give significant contributions to the behavior in the stable phase, including an increased critical damage when compared with ELS.

What happens in more realistic scenarios with intermediate interaction ranges, like in e.g.\ the $\gamma$-model \cite{hmkh02} or soft clamp model \cite{bhs02}, is still an open question. One could speculate that such models should be somewhere between the ELS and LLS models, and that they might contain a weaker shielding effect than in the LLS model, but a thorough analysis would be required to give definite answers.

\begin{acknowledgments}
We would like to thank Martin Hendrick for insightful discussions. This work was partly supported by the Research Council of Norway through its Centers of Excellence funding scheme, project number 262644. EB was funded by the Research Council of Norway through project number 250158.
\end{acknowledgments}


\end{document}